# A ground-based near-infrared emission spectrum of the exoplanet HD 189733b

Mark R. Swain[1], Pieter Deroo[1], Caitlin A. Griffith[2], Giovanna Tinetti[3], Azam Thatte[5], Gautam Vasisht[1], Pin Chen[1], Jeroen Bouwman[6], Ian J. Crossfield[4], Daniel Angerhausen[7], Cristina Afonso[6,] & Thomas Henning[6]

*[1]Jet Propulsion Laboratory, California Institute of Technology, 4800 Oak Grove Drive, Pasadena, California 91109-8099 USA. [2]University of Arizona, Lunar and Planetary Laboratory, Space Science Bldg. Room 525, 1629 E. University Blvd. Tucson, AZ 85721 USA [3]Department of Physics and Astronomy, University College London, Gower Street, London WC1E 6BT. [4]Astronomy Department UCLA, 475 Portola Plaza, Los Angeles, CA 90034 USA. [5]Woodruff School of Mechanical Engineering, MRDC Building, Room 4111,Georgia Institute of Technology, Atlanta, GA 30332-0405 USA. [6] Max-Planck Institute for Astronomy, Koenigstuhl 17, D-69117 Heidelberg, Germany. [7]German SOFIA Institute, Institute for Space Systems, Pfaffenwaldring 3170569 Stuttgart, Germany*

**Detection of molecules via infrared spectroscopy probes the conditions and composition of exoplanet atmospheres. To date, water ($H_2O$), methane ($CH_4$), carbon dioxide ($CO_2$), and carbon monoxide (CO) have been detected[1-5] in two hot-Jupiter type exoplanets. These previous results relied on space-based telescopes that do not provide spectroscopic capability in the important 2.4–5.2 µm spectral region. Here we report ground-based observations of the dayside emission spectrum for HD 189733b between 2.0–2.4 µm and 3.1–4.1 µm, where we find a bright emission feature. Where overlap with space-based instruments exists, our results are in excellent agreement with previous measurements[2,6]. The feature around 3.25 µm is completely unexpected and is difficult to explain with models that assume local thermodynamic equilibrium**

**(LTE) conditions in the 1 bar to 1×10$^{-6}$ pressures typically sampled by infrared measurements. The most likely explanation for this emission is non-LTE emission from $CH_4$, similar to what is seen in the atmosphere of planets in our own solar system[7-9]. These results suggest that non-LTE effects may need to be considered when interpreting measurements of strongly irradiated exoplanets.**

Spectroscopic detection of molecules in exoplanet atmospheres is a relatively recent development. High-precision spectroscopy data in combination with spectral retrieval techniques enables the determination of the atmospheric temperature profile and composition; this, in turn, provides some degree of knowledge about the atmospheric chemistry via molecular abundance ratios. Combinations of the molecules $H_2O$, $CH_4$, $CO_2$, and CO have been detected in the hot-Jupiter-type exoplanets HD 189733b and HD 209458b during primary eclipse[1,2] and/or secondary eclipse[3-5] events (when the planet passes either in front of or behind the stellar primary). In the case of HD 189733b, primary and secondary eclipse spectra, respectively, probe the dayside and terminator regions of the planet's atmosphere and thus permit atmospheric conditions in these two regions to be compared. Temporal variability has also been detected in the mid-infrared emission spectrum[3] and may indicate that dynamical processes have a significant role in establishing the atmospheric properties[10,11]. To date, the detection of molecules in exoplanet atmospheres has required the stability of the Hubble and Spitzer space telescopes. Observing exoplanet emission spectra from the ground is complicated by both the intrinsic variability of the Earth's atmosphere and instrument instability caused, for example, by changes in the instrument orientation as the telescope tracks.

The spectrum we present in this paper is based on data obtained on August 11th, 2007, using data from the SpeX instrument[12] on the NASA Infrared Telescope Facility (IRTF). The observations of the HD 189733 system were timed to observe the secondary eclipse light curve, beginning approximately one hour before the onset of ingress and ending approximately one hour after the termination of egress. HD 189733 was observed at two positions (A and B) on the spectrograph entrance slit using an AB BA sequence. Spectral calibration was done using the internal instrument calibration arc lamps. To minimize the effects of seeing-based modulation of the measured flux density, a 1.6 arc second slit width was used. The SpeX instrument was configured to observe between 1.9–4.2 μm at an average spectral resolution of 470. The slit length for these observations was 15 arcseconds, and the sky background and source were measured simultaneously.

We have developed a novel calibration method (described in the Supplementary Information) for systematic error removal and subsequent detection of the eclipse. This method is based on an iterative approach to removing systematic errors while the secondary eclipse is extracted by computing the self-coherent spectrum of groups of channels[13]. Although we have not reached the theoretical noise level and further improvements are likely possible, this approach is highly successful in detecting the secondary eclipse and, significantly, does not require the use of priors (e.g. system ephemeris). We validated the calibration/spectral extraction method by comparing the ground-based results to previous space-based measurements of HD 189733b. Between 2.0–2.4 μm, ground-based results clearly show the $CO_2$ absorption feature present in the exoplanet atmosphere and agree well with previous Hubble measurements[2] (see Fig. 1). In the 3.1–4.1μm region, we compared the ground-based results with Spitzer photometry measurements[6] by integrating the ground-based spectrum over the Spitzer 3.6 μm filter pass band, and we find agreement at the 1-σ level. Between 3.1–4.1 μm, the ground-based observations provide a unique capability, and it is here that we find emission features far exceeding those characteristic of other wavelengths in the HD 189733b dayside spectrum (see Fig. 2).

We investigated the plausibility of purely thermal emission by calculating the brightness temperature spectrum; these values indicate the atmospheric temperatures probed by the light emitted at each wavelength when LTE conditions (collision dominated) apply. The 3.25 μm flux density corresponds to a brightness temperature of 2700 K, which markedly exceeds the typical brightness temperatures in HD189733's spectrum (see Fig. 2 inset). Prior studies[3,6,11,14] and our own modelling successfully explained most of the infrared spectrum emission as thermal emission modulated by molecular opacity; these studies imply a temperature profile that decreases from roughly 1600 K at 1 bar to roughly 800 K in the $\sim 10^{-3}$–$10^{-5}$ bar range. However, these LTE models do not explain the 3.1–4.1 μm flux density; modification of these models to include a tropopause and an appropriate absorber could explain the weak emission around 3.8 μm. Applying LTE models to explain the 3.25 μm emission requires a temperature structure and $CH_4$ abundance that is incompatible with all other portions of the dayside spectrum; the inclusion of a "hot layer" of $CH_4$ has difficulty explaining the spectrum without causing the brightness temperature to increase around 2.2 and 7.8 μm (the Octad and $\nu_4$ band respectively). Thus, we conclude that the dayside emission at between 3.1–3.4 μm is probably caused by a non-LTE emission process.

In our own solar system, $CH_4$ fluorescence has been detected in Jupiter[7,8], Saturn[7,8], and Titan[9]. The 3.1 and 3.25 μm features in the HD 189733b dayside spectrum resemble those of Titan, where the fluorescence of the $\nu_3$ bands of $CH_4$ and HCN at pressures below $10^{-4}$ bar cause emission far exceeding that possible in LTE conditions. The absence of a 3.9 μm feature suggests that $H_3^+$ emission is not significant in HD 189733b. When taken together, (1) the approximate alignment between the HD 189733b spectrum and the peaks in the observed fluorescence spectrum of Titan, (2) the difficulty of explaining the 3.1–3.4 μm band emission in HD 189733b using LTE models, and (3) the prevalence of fluorescence in our own solar system, indicate that non-LTE $CH_4$ emission is likely present in the atmosphere of HD 189733b. A puzzling aspect of these data is the absence of strong emission from the $CH_4$ $\nu_3$ band P branch; a similar effect has been detected in the upper atmosphere of Titan (see Supplementary Information) and remains unexplained.

The 3.1–4.1 μm spectrum shows that the strong emission from the $CH_4$ $\nu_3$ band dominates the 3.6 μm Spitzer photometry measurement of this planet (see Fig. 1). This is significant because previous models, based on the assumption of LTE, systematically underpredict the 3.6 μm Spitzer flux density measurement[3,6,11,14]. The underprediction is now easily explained by the presence of strong non-LTE emission in the filter pass band. Non-LTE emission may also explain another puzzling aspect of HD 189733b in which the onset of secondary eclipse in the Spitzer 3.6 μm measurements is delayed by 5.6 ± 0.8 minutes relative to other Spitzer photometric bands[6]; the authors speculated that this delay "may arise from strong brightness variations across the visible hemisphere of the planet". Given that the $CH_4$ $\nu_3$ band lies within the Spitzer 3.6 μm filter band pass and that emission at other wavelengths in the filter band pass is weak (see Fig. 1), a significant non-uniformity of the non-LTE emission in HD 189733b could produce the observed delay in the onset of secondary eclipse. Possible sources for spatially inhomogeneous non-LTE emission include localized vertical transport of $CH_4$ from lower levels of the atmosphere or the presence of clouds.

Of equal significance to the discovery of non-LTE emission in an exoplanet atmosphere is the ground-based detection of molecules in an exoplanet atmosphere. We believe the calibration method used here can be applied directly to many existing instruments. The IRTF is not considered a "large" telescope, and the SpeX instrument has no specific optimization for high-dynamic-range spectroscopy; this suggests that many facilities could make similar measurements and that improvements could be realized by purpose-built instrumentation on large telescopes. Given the availability of numerous large telescopes equipped with infrared spectrometers, the result in this paper foreshadows a large quantity of "molecular-abundance-grade" exoplanet spectra. Although telluric opacity limits ground-based observations to certain spectral regions, these results, together with recent optical detections of atomic features[15,16], decisively show that ground-based spectroscopy will have a powerful and lasting influence in the emerging field of exoplanet characterization.

Hot-Jupiter-type planets experience powerful radiation forcing due to close proximity to the stellar primary and, in the cases of HD 189733b and HD 209458b, are thought to be tidally locked as well. No analog for these planets exists in our own solar system, and thus this class of object represents an opportunity to study planetary atmospheres in a completely new regime. Detection of molecules via infrared spectroscopy is currently probing the conditions, composition, and chemistry of HD 189733b and HD 209458b and, by extension, other members of the hot-Jupiter exoplanet class. Previous observations have shown both relatively high $CO_2$ & $CH_4$ abundances[4,5] and temporal variability[3]; the former indicates the possible role of non-equilibrium chemistry, while the latter indicates the possible presence of dynamical effects. The results in this paper show that interpreting hot-Jupiter spectra by assuming only LTE conditions is questionable. Thus, the simple picture of a time-stationary exoplanet atmosphere governed by LTE conditions and dominated by equilibrium chemistry is being challenged. The observations reported here are the latest in a series of results supporting the view that the highly irradiated gas giant planets exhibit a degree of complexity, together with a rich variety of physical and chemical processes, that we have only begun to understand

**Acknowledgements** We thank Bobby Bus at the IRTF for several helpful discussions regarding the operation of the SpeX instrument and for his support during our observing runs. We thank the observing staff at the IRTF for their assistance and advice during observing runs. We thank Linda Brown for making recommendations on molecular line lists and Glen Orton for extensive discussions about the interpretation of these results. G. Tinetti was supported by the UK Sciences & Technology Facilities Council and the European

Space Agency. The research described in this paper was carried out at the Jet Propulsion Laboratory, California Institute of Technology, under a contract with the National Aeronautics and Space Administration.

**Contributions** M. Swain wrote the paper, participated in developing the calibration methods, and wrote the telescope proposals. P. Deroo worked on developing and testing the calibration method. C. Griffith and G. Tinetti worked on the interpretation of the results. A. Thatte worked on the calibration method. G. Vasisht and P. Chen worked on calibration validation. J. Bouwman worked on the calibration of the Spitzer data. I. Crossfield worked on an early version of the calibration method. D. Angerhausen contributed to the atmospheric calibration approach. C. Afonso and T. Henning were co-authors on the original telescope proposal and provided comments on the paper text.

**Author Information** Reprints and permissions information is available at www.nature.com/reprints. The authors declare no competing interests. Correspondence and requests for materials should be addressed to M. Swain (e-mail: Mark.R.Swain@jpl.nasa.gov).

(Fig. 1) **Dayside spectra and secondary eclipse light curves.** The ground-based spectra are in excellent agreement with space-based measurements. The upper panel shows the comparison of the IRTF measurements (red) with previous measurements using the Hubble Space Telescope (green). The lower panel shows the comparison of the IRTF spectrum (red) and IRTF data averaged to the Spitzer 3.6 μm pass band (blue solid square) with the Spitzer Space Telescope 3.6 μm photometry measurement (blue open square). Each calibrated light curve includes the averaged measurements (black diamonds with ± 1-σ errors), and the best-fit eclipse model (red). In the comparison between ground-based and space-based measurements, the secondary eclipse depth is shown as the planet/star contrast ratio.

(Fig. 2) **Unexpectedly strong 3.25 μm emission present in the dayside spectrum.** The brightness temperature of the 3.25 μm emission feature indicates the likely presence of a non-LTE emission mechanism. The dayside emission spectrum is based on the new measurements reported in this paper (black), together with previous results from Hubble spectroscopy (red), Spitzer spectroscopy (green), and Spitzer photometry (blue); all data are shown with ± 1-σ errors. A radiative transfer model (grey) assuming LTE conditions and consistent with the measurements made with Spitzer and Hubble fails to describe the emission structure at 3.1–4.1 μm, and we find no plausible combination of atmospheric parameters that provides a good model of the observations under LTE conditions. The inset plot displays the brightness temperature at each wavelength and shows the large temperature change needed to produce the 3.25 μm emission if LTE conditions hold.

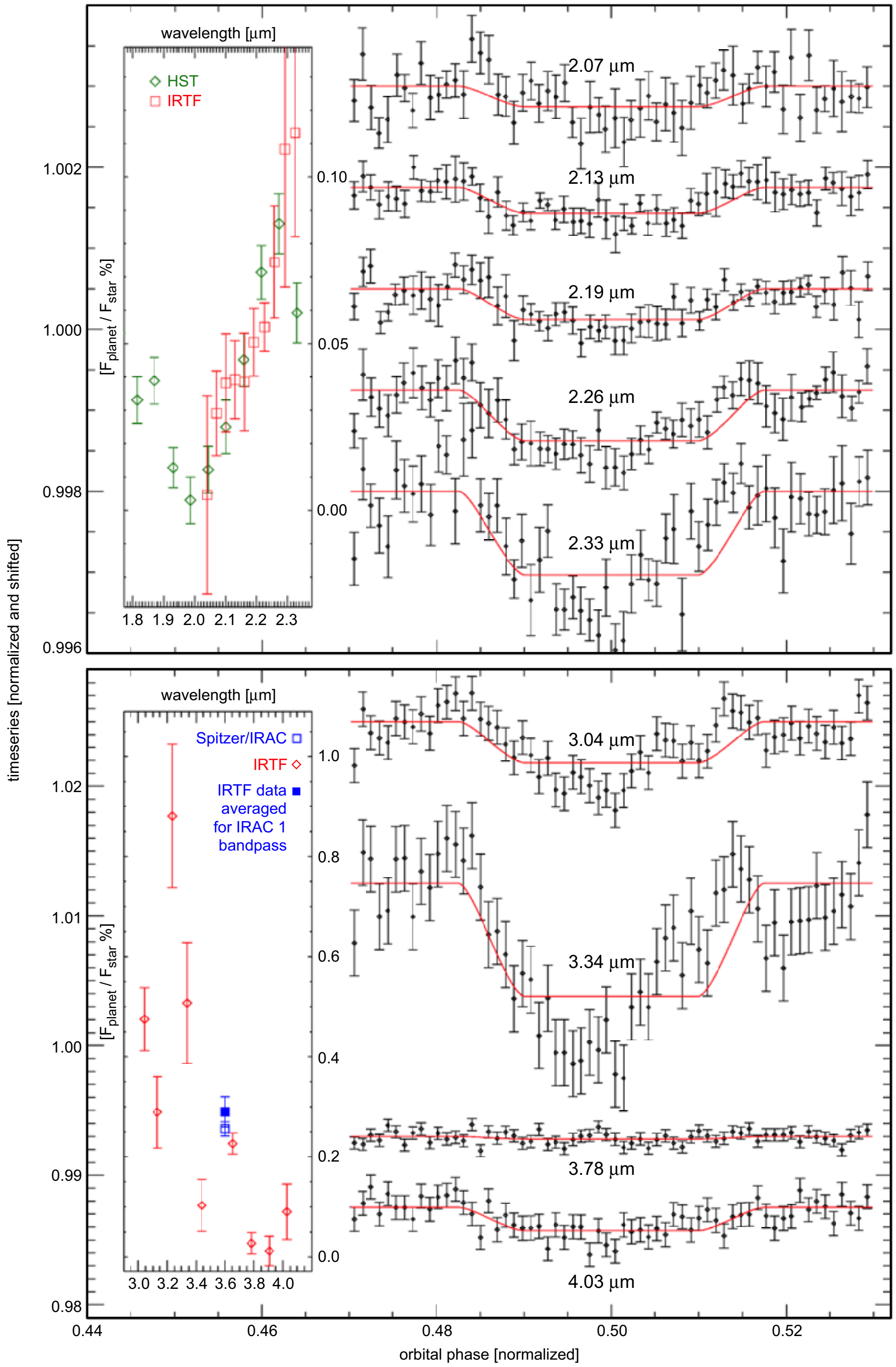

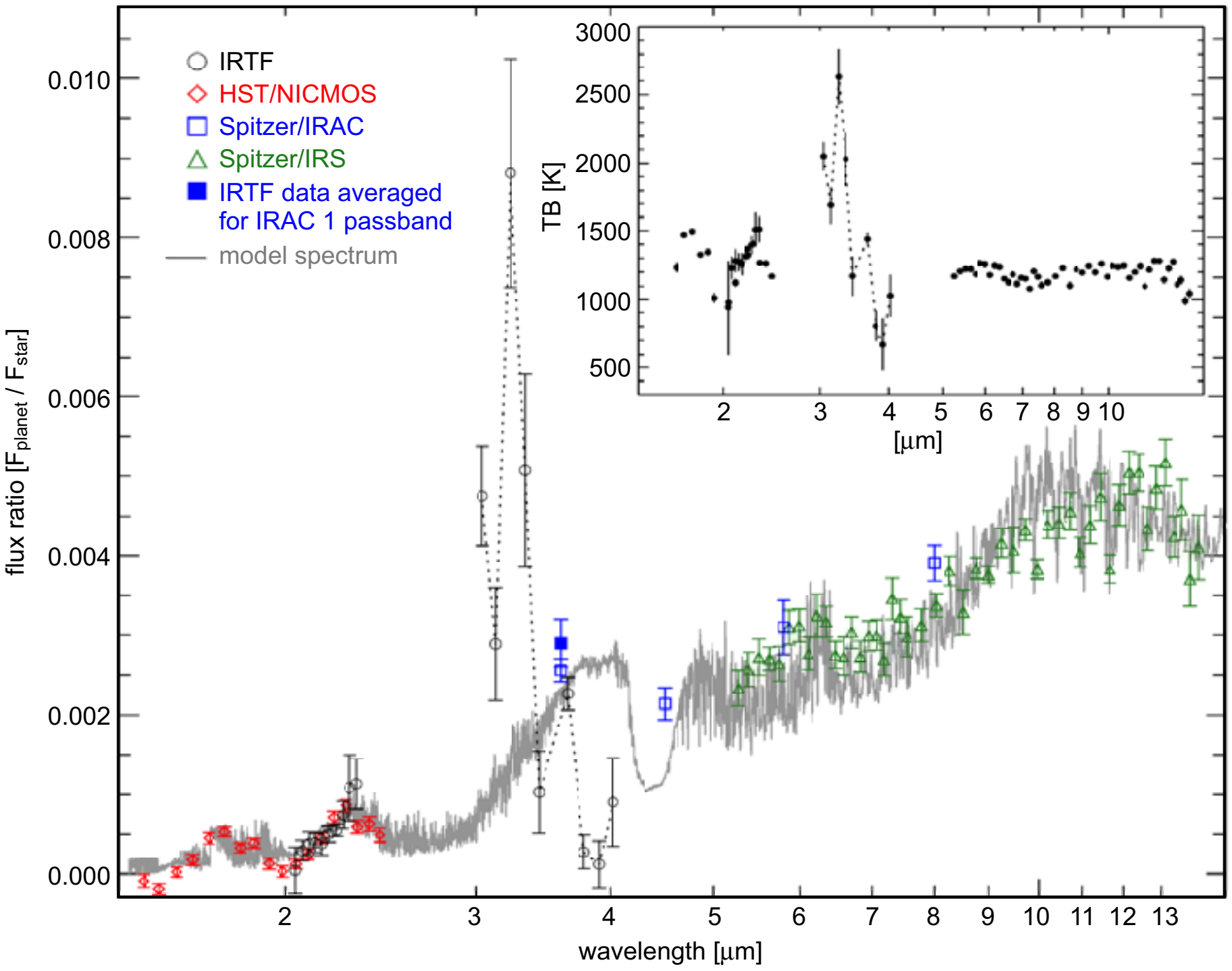